\documentclass[twoside]{articlek}
\usepackage{float}

\renewcommand{\refname}{REFERENCES}
\def\be{\begin{equation}}
\def\ee{\end{equation}}

\def\BibTeX{{\rm B\kern-.05em{\sc i\kern-.025em b}\kern-.08em
            T\kern-.1667em\lower.7ex\hbox{E}\kern-.125emX}}

\usepackage{graphicx}
\begin{document}
\sloppy
\twocolumn[{
{\large\bf Open-charm and charm-tagged-jet production with ALICE at LHC}\\

{\small J. Kvapil, jakub.kvapil@cern.ch, School of Physics and Astronomy, the University of Birmingham, Edgbaston, Birmingham, UK. B15 2TT}\\

{\small \textbf{ABSTRACT} The latest results on charm-tagged jets and open-charm production obtained with \mbox{ALICE} at the LHC are presented. The baryon-to-meson ratios for $\mathrm{\Lambda^+_c}$, $\mathrm{\Sigma^{0,+,++}_c}$ and $\mathrm{\Xi^{0,+}_c}$ measured in pp collisions at $\sqrt{s}=13\ \mathrm{TeV}$ are discussed. The measurements of the jet-momentum fraction carried by the $\mathrm{D^0}$ meson and $\mathrm{\Lambda^+_c}$ baryon in pp collisions at $\sqrt{s}=13$~TeV are reported, and the number of leading-parton splittings $n_{SD}$ for $\mathrm{D^0}$-tagged and inclusive jets in pp collisions at $\sqrt{s}=13$~TeV are compared. Finally, the first direct measurement of the dead-cone effect at hadron colliders is shown. }

\vspace*{2ex}

}]

\section{INTRODUCTION}
Charm quarks are mostly produced in hard partonic scattering processes in the early stages of
hadronic collisions. Because of their large mass, their production cross section can be calculated using perturbative quantum chromodynamics down to zero transverse momentum $p_{\mathrm{T}}$. Moreover, the study of the charm baryon-to-meson ratio provides unique information on the hadronisation mechanism. Measurements of charm-tagged jets add further information on the charm-quark fragmentation. 



\section{RESULTS} 
\begin{figure} [ht!]                     
	\begin{center}                        
		\includegraphics[width=52mm]{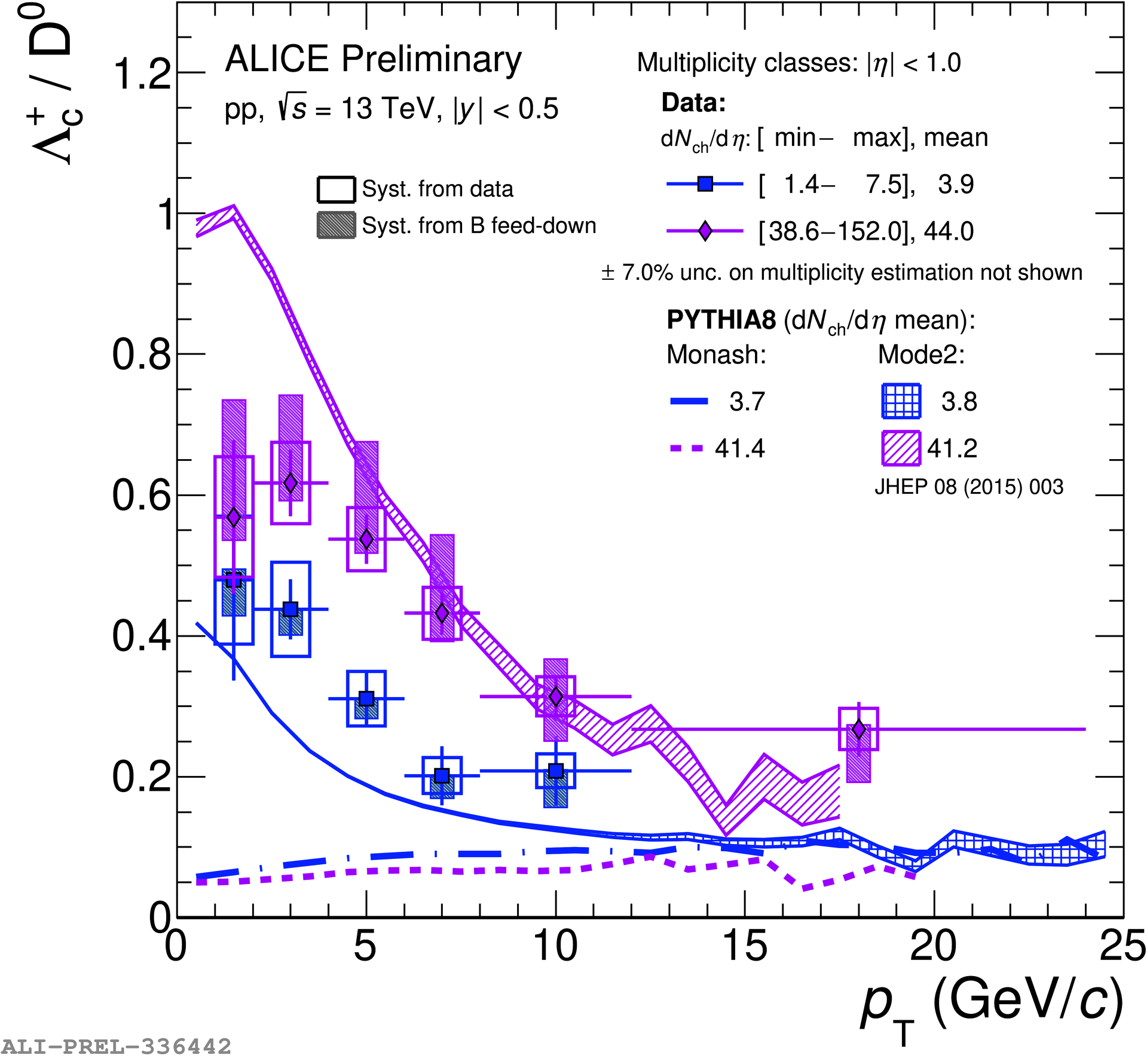}   
	\end{center}                         
	\vspace{-2mm} \caption{The $p_{\mathrm{T}}$-differential $\mathrm{\Lambda^+_c/D^0}$ ratio for two multiplicity intervals in pp collisions at $\sqrt{s}=13$ TeV compared to the default PYTHIA8 tune \cite{P8,LU} and to a tune with colour-reconnection beyond leading colour approximation \cite{CR}.}
	\label{lcratio}
\end{figure}

The ratio of the $p_{\mathrm{T}}$-differential cross sections of $\mathrm{\Lambda^+_c}$ baryon and $\mathrm{D^0}$ meson in pp collisions at $\sqrt{s}=13$~TeV is shown in Fig. \ref{lcratio} for two intervals of charged-particle multiplicity. The data show strong $p_{\mathrm{T}}$ and multiplicity dependence and are described within uncertainties by a tune of PYTHIA8 with colour reconnections beyond the leading colour approximation \cite{P8,CR}. On the other hand, the default version of PYTHIA8 \cite{LU}, which is tuned to reproduce data from $\mathrm{e^+e^-}$ collisions, underestimate the ratios and does not reproduce their multiplicity dependence. 


\begin{figure} [t!]                     
	\begin{center}                        
		\includegraphics[width=55mm]{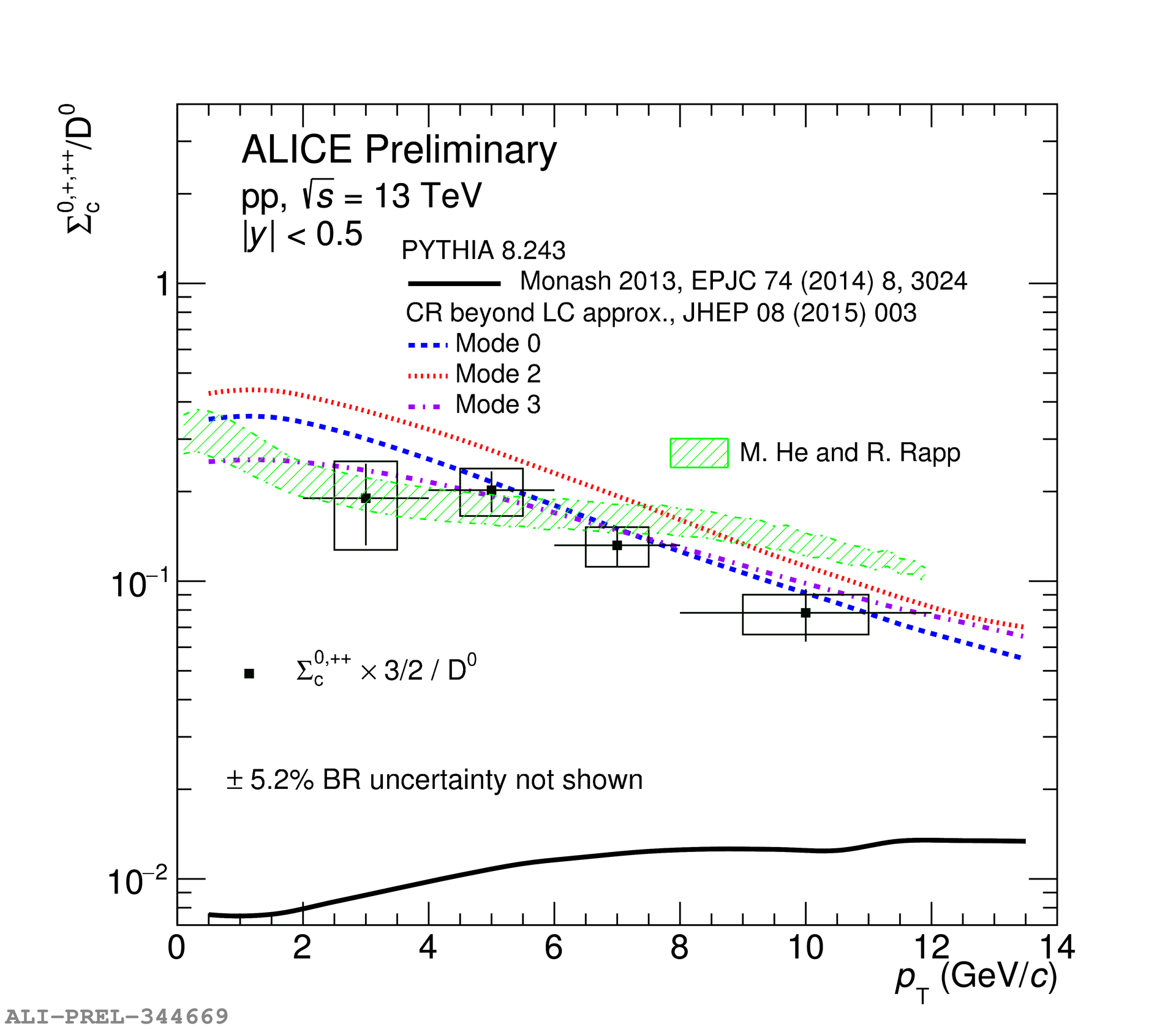}   
	\end{center}                         
	\vspace{-2mm} \caption{The $p_{\mathrm{T}}$-differential $\mathrm{\Sigma^{0,+,++}_c/D^0}$ ratio in pp collisions at $\sqrt{s}=13$ TeV compared to the default~PYTHIA8 tune \cite{P8,LU} and to a tune with colour-reconnection beyond leading colour approximation~\cite{CR}.}
	\label{sigratio}
\end{figure}

\begin{figure} [t!]                     
	\begin{center}                        
		\includegraphics[width=52mm]{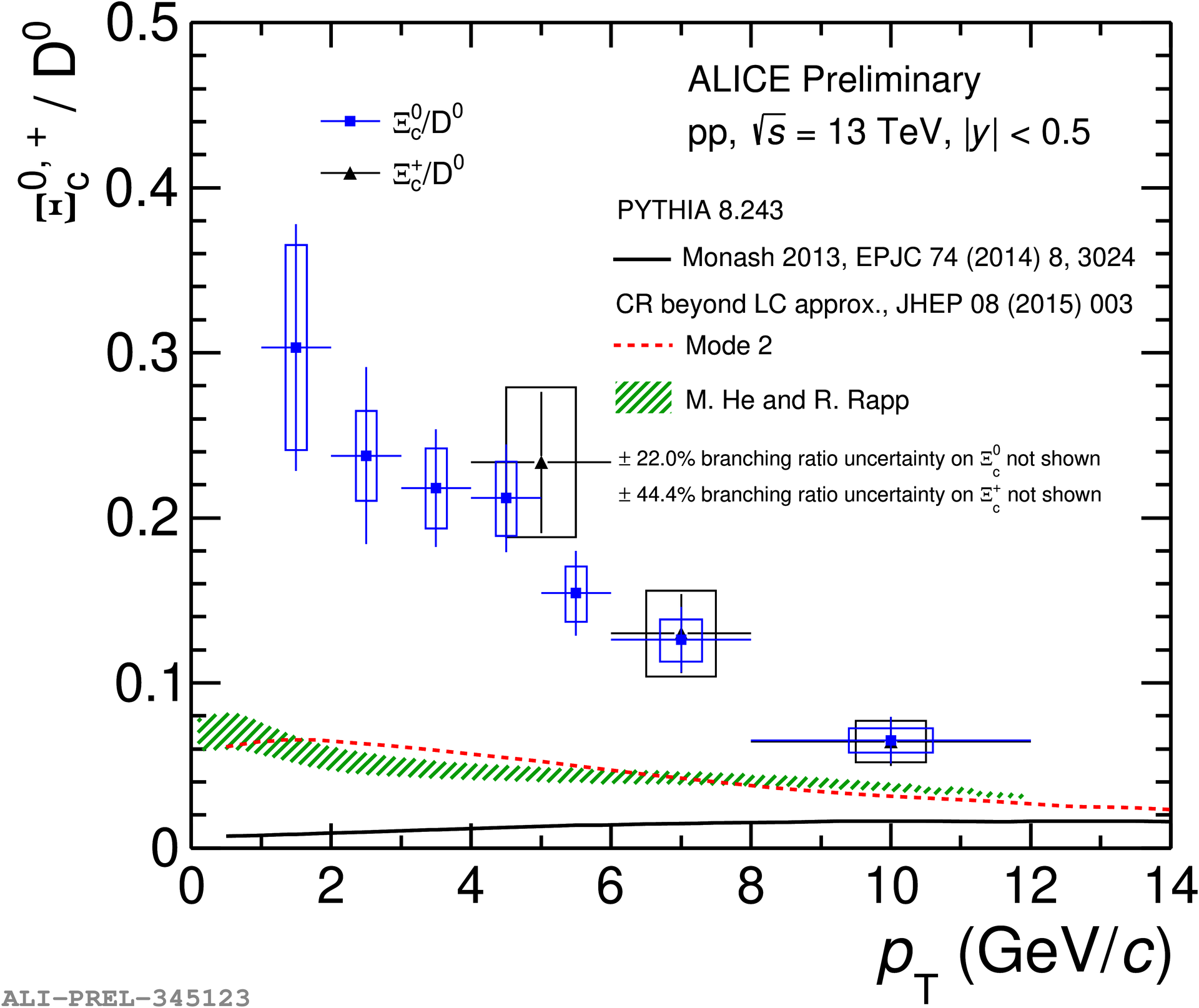}   
	\end{center}                         
	\vspace{-2mm} \caption{The $p_{\mathrm{T}}$-differential $\mathrm{\Xi^{0,+}_c/D^0}$ ratio in pp collisions at $\sqrt{s}=13$ TeV compared to the default PYTHIA8 tune \cite{P8,LU} and to a tune with colour-reconnection beyond leading colour approximation \cite{CR}.}
	\label{xiratio}
\end{figure}

The baryon-to-meson ratios $\mathrm{\Sigma^{0,+,++}_c/D^0}$ and $\mathrm{\Xi^{0,+}_c/D^0}$ are shown in Fig. \ref{sigratio} and \ref{xiratio}, respectively. The PYTHIA8 tune with colour reconnection beyond leading-colour approximation describes the $\mathrm{\Sigma^{0,+,++}_c/D^0}$ ratio, while it underestimates the $\mathrm{\Xi^{0,+}_c/D^0}$ one.

\begin{figure} [t!]                     
	\begin{center}                        
		\includegraphics[width=47mm]{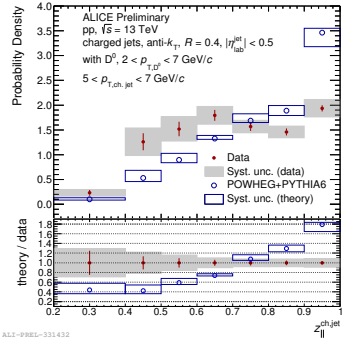}   
	\end{center}                         
	\vspace{-2mm} \caption{Probability density distribution of the jet momentum fraction, $z^{ch}_{\parallel}$, carried by $\mathrm{D^0}$ mesons measured in pp collisions at $\sqrt{s}=13\ \mathrm{TeV}$ for $5<p_{\mathrm{T,jet}}<7\ \mathrm{GeV/}c$ compared to POWHEG+PYTHIA6 predictions \cite{P6,PO}.}
	\label{fig:D}
\end{figure}

\begin{figure} [t!]                     
\begin{center}                        
	\includegraphics[width=55mm]{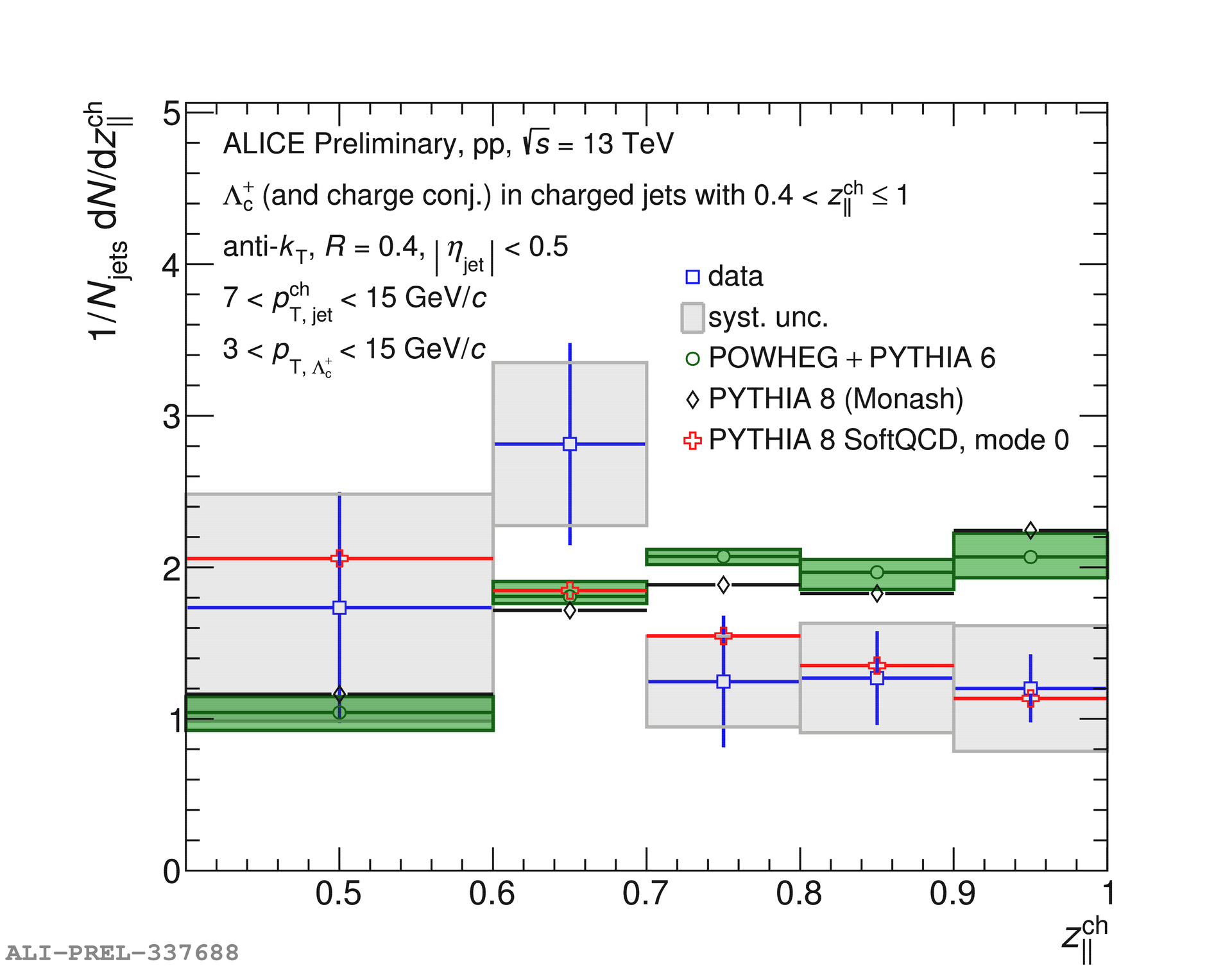}   
\end{center}                         
\vspace{-2mm} \caption{Probability density distribution of the jet momentum fraction, $z^{ch}_{\parallel}$, carried by $\mathrm{\Lambda^+_c}$ baryon measured in pp collisions at $\sqrt{s}=13\ \mathrm{TeV}$ for $7<p_{\mathrm{T,jet}}<15\ \mathrm{GeV/}c$ compared to expectations from POWHEG+PYTHIA6 \cite{P6,PO} and PYTHIA8 \cite{P8,CR,LU}.}
\label{fig:D2}
\end{figure}

Jets are reconstructed using the anti-$k_\mathrm{T}$ algorithm and are tagged as $\mathrm{D^0}$-meson or $\mathrm{\Lambda^+_c}$-baryon jets if they contain a fully reconstructed $\mathrm{D^0}$ meson or $\mathrm{\Lambda^+_c}$ baryon. The parallel jet momentum fraction, $z^{ch}_{\parallel}=\frac{\vec{p}_\mathrm{D}\cdot \vec{p}_{\mathrm{ch. jet}}}{\vec{p}_\mathrm{ch. jet}\cdot \vec{p}_\mathrm{ch. jet}}$, is proportional to the emitted angle of the heavy-flavour particle with respect to the jet axis. The $z^{ch}_{\parallel}$ probability densities of $\mathrm{D^0}$-tagged jets with $5<p_{\mathrm{T,jet}}<7\ \mathrm{GeV/}c$ and $\mathrm{\Lambda^+_c}$-tagged jets with $7<p_{\mathrm{T,jet}}<15\ \mathrm{GeV/}c$ for pp collisions at $\sqrt{s}=13\ \mathrm{TeV}$ are shown in Fig. \ref{fig:D} and \ref{fig:D2}, respectively. The POWHEG+PYTHIA6 model predicts harder fragmentation than the measured ones. In addition, the data favour models with colour reconnections beyond leading colour approximation.

\begin{figure} [t!]                     
	\begin{center}                        
		\includegraphics[width=50mm]{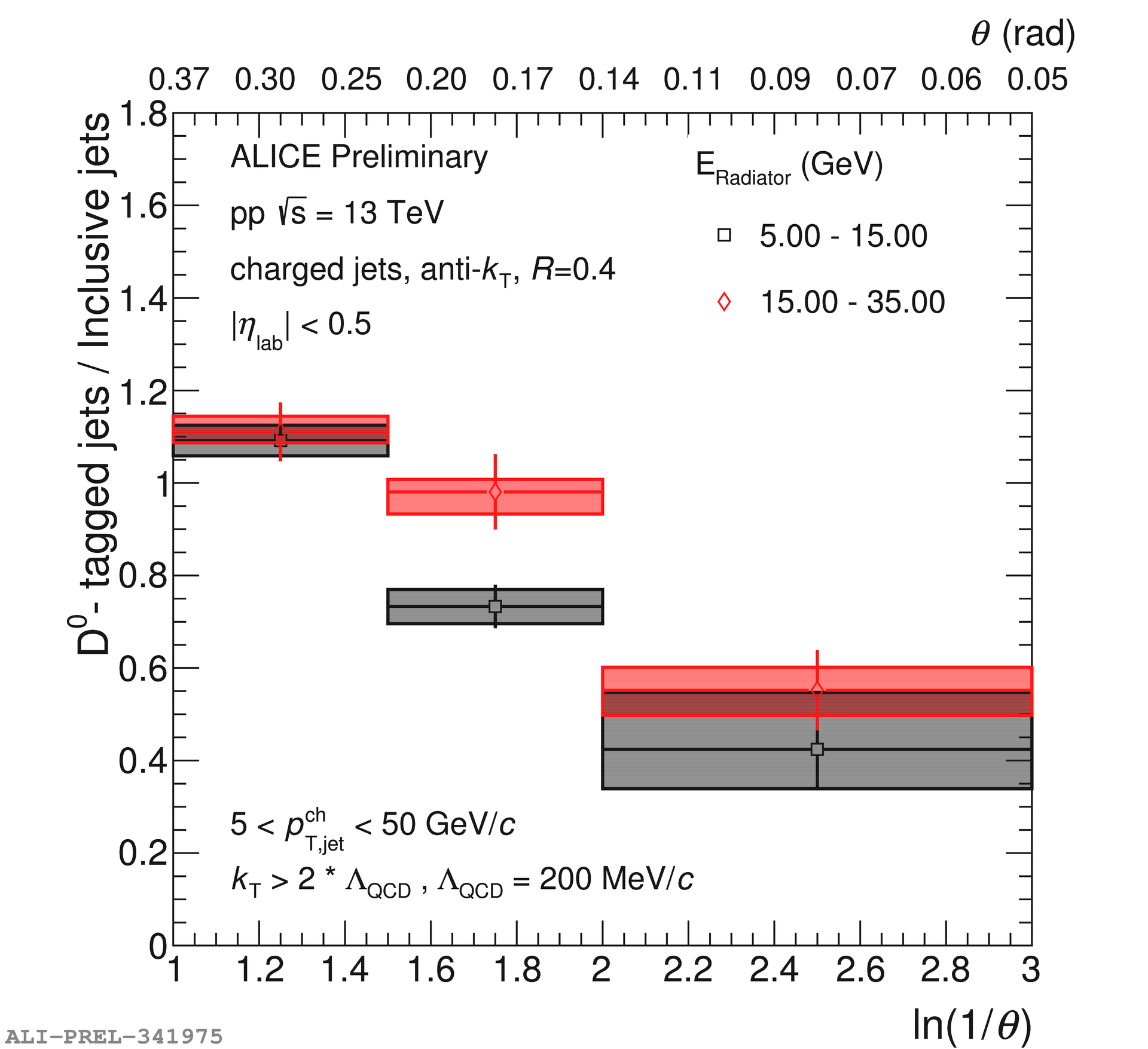}   
	\end{center}                         
	\vspace{-2mm} \caption{Ratio of the splitting-angle distributions of $\mathrm{D^0}$-tagged jets and inclusive jets in pp collisions at $\sqrt{s}=13\ \mathrm{TeV}$.}
	\label{deadcone}
\end{figure}

\begin{figure} [t!]                     
	\begin{center}                        
		\includegraphics[width=50mm]{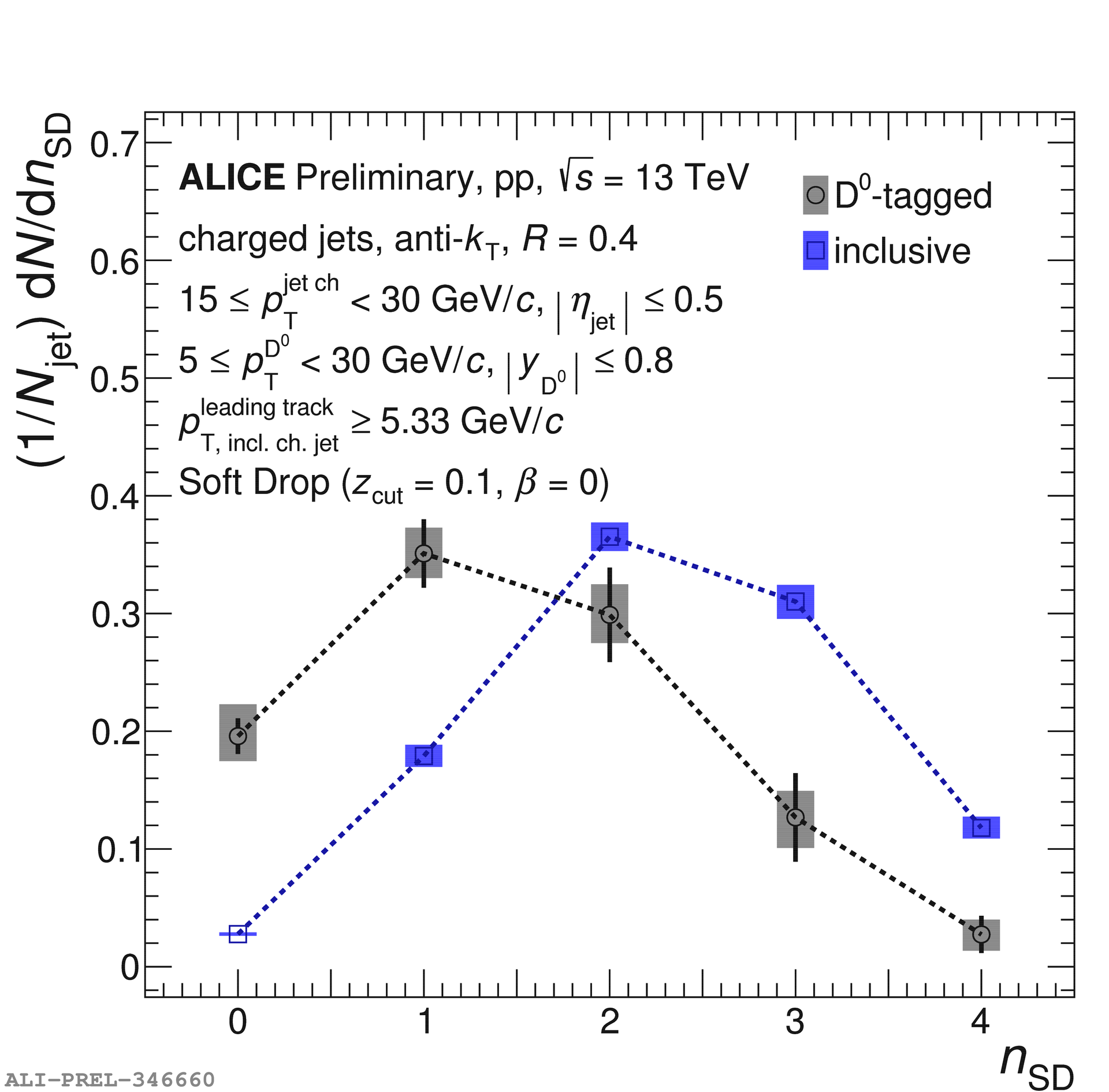}   
	\end{center}                         
	\vspace{-2mm} \caption{Number of leading-parton splittings $n_{SD}$ in pp collisions at $\sqrt{s}=13$ TeV for $\mathrm{D^0}$-meson tagged and inclusive jets.}
	\label{nsd}
\end{figure}

One of the fundamental QCD effect is the so-called ``dead-cone" \cite{DC}, which suppresses gluon radiation at small angles with respect to the quark direction for massive quarks. Jets are declustered to access the hard partonic splitting and the ratio of splittings angle distribution for $\mathrm{D^0}$-tagged and inclusive jets is reported in Fig. \ref{deadcone}. Splittings with small angle are suppressed for $\mathrm{D^0}$-tagged jets with respect to inclusive jets. This is the first direct evidence of dead-cone at hadron colliders. Another consequence of the dead-cone effect, visible in Fig. \ref{nsd}, is the smaller number of leading-parton splittings $n_{SD}$ for $\mathrm{D^0}$-tagged jets with respect to inclusive jets. 

\section{CONCLUSION} 
The hadronization of charm quarks is different in $\mathrm{e^+e^-}$ and in pp collisions and, in the latter, depends on the multiplicity of charged particles produced in the event. A tune of PYTHIA8 with colour reconnection beyond leading-colour approximation describes well the $\mathrm{\Lambda^+_c/D^0}$ and $\mathrm{\Sigma^{0,+,++}_c/D^0}$ baryon-to-meson ratios, while it fails to reproduce the $\mathrm{\Xi^{0,+}_c/D^0}$ one. In general, models predict harder fragmentation of charm quark in jets than what observed in data. Finally, the charm quark in jets splits less in comparison to inclusive jets mainly because of the dead-cone effect.

\leftskip=-5pt \vspace{-0.3truecm}
\bibliographystyle{ieeetr}
\bibliography{CiteDatabase}




\end{document}